\documentstyle[proceedings,psfig]{crckapb}
\begin{opening}
\title{Shock Waves of the Large-Scale Structure Formation in the Universe}
\author{T.A. Ensslin}
\author{P.L. Biermann}
\institute{Max--Planck--Institut f\"ur Radioastronomie Bonn, Germany}
\author{U. Klein}
\author{S. Kohle}
\institute{Radioastronomisches Institut der Universit{\"a}t Bonn, Germany}
\end{opening}

\begin{document}

\begin{abstract}
Simulations of structure formation in the Universe predict accretion
shock waves at the boundaries of the large-scale structures as sheets,
filaments, and clusters of galaxies. If magnetic fields are present at
these shocks, particle acceleration should take place, and could
contribute to the observed cosmic rays of high energies. When the
radio plasma of an old invisible radio lobe is dragged into such a
shock wave, the relativistic electron population will be reaccelerated
and the plasma becomes radio-luminous again. Such tracers of the
accretion shock waves are observed at the boundaries of some clusters
of galaxies: the so-called cluster radio relics, which are large
regions of diffuse radio emission, without any parent galaxy
nearby. The observed properties of the cluster radio relics are
naturally explained by accretion shock waves. Radio relics therefore
give the first evidence for the existence of accretion shocks of the
large-scale structure formation and they allow investigations of the
shock properties.
\end{abstract}
\section{ Shock Waves of the Large-Scale Structure Formation }
The large-scale structure of the Universe, seen in the
structured galaxy distribution, is still forming. Matter is flowing
out of the cosmic voids onto sheets and filaments. Within the
filaments the matter flows to the density cusps frequently located at
the intersection points of filaments: the clusters of
galaxies. Whenever the flow passes from one structure into another,
its velocity suddenly changes and several Mpc sized shock waves
occur. At these shock waves the kinetic energy of the gravitationally
accelerated gas is dissipated, mainly thermalized to temperatures of
a few $0.1$ keV in filaments and several keV in clusters of
galaxies. The shock velocity at filaments is expected to be of the
order of several $100$ km s$^{-1}$, and the accretion shocks at
clusters $1000-2000$ km s$^{-1}$. From simulation of structure
formation rough values of the cluster accretion shock radius and
velocity can be given in terms of the cluster temperature as an
indicator of the gravitational potential (Kang et al. 1997):
\begin{eqnarray}
\label{eq:rs}
r_{\rm s} &=& 4.24\, h_{50}^{-1} {\rm Mpc} \,( kT_{\rm
obs}/{6.06\,{\rm keV}} )^{1/2} (1+z)^{-3/2}\\
\label{eq:Vs}
V_{\rm s, predicted} &=& 1750 \, {\rm km\, s^{-1}}\, ({kT_{\rm
obs}}/{6.06\,{\rm keV}} )^{1/2}\,\,.
\end{eqnarray}
The dissipated accretion power per shock surface is:
\begin{equation}
\label{eq:Qflow}
Q_{\rm flow} \approx \frac{1}{2}\,n_{\rm e}\,m_{\rm p}\,V_{\rm s}^3
\approx 4\cdot 10^{44} \;\, \frac{\rm erg\,s^{-1}}{\rm Mpc^2}\,\,\frac{n_{\rm 
e}}{10^{-5}\,{\rm cm^{-3}}}\,\left( \frac{kT_{\rm obs}}{6.06\,{\rm keV}}
\right)^{3/2}\,.
\end{equation}
\section{ Particle Acceleration at Cluster Accretion Shocks}
If magnetic fields are present at the location of the accretion shocks,
charged particles are accelerated. The acceleration of protons is mainly
limited by photo-meson production to $< 10^{20}$ eV, but could contribute to
the UHECR spectrum below this energy (Kang et al. 1997). Relativistic
electrons lose their energy by IC scattering of MWB photons and synchrotron
losses. Especially the synchrotron radio emission should be a sensitive
tracer of shock waves. The best detectability is given if a strongly magnetized
plasma is present with a preaccelerated population of electrons. This is the
case whenever an old remnant radio lobe of a former radio galaxy is dragged
into such a shock.  Since the Universe was sufficiently polluted by radio
plasma at the epoch of quasar activity (En{\ss}lin et al. 1997, 1998b),
such shocks should be traceable today. We argue that this is the case for
cluster radio relics.

\section{ Cluster Radio Relics}
\begin{figure}
\psfig{figure=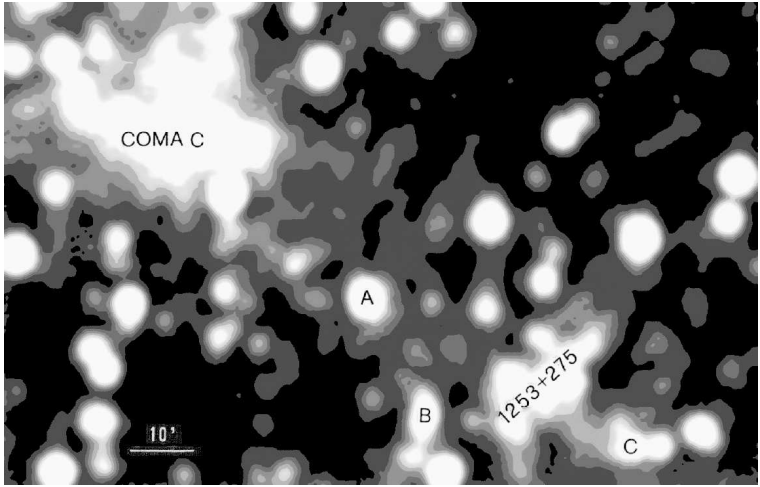,width=0.8\textwidth}
\caption[]{\label{fig:coma}Westerbork Synthesis Radio Telescope map at
327 MHz of the Coma cluster from Giovannini et al. (1991). The central
halo source Coma C and the cluster relic 1253+275 are
identified. Capital letters indicate some extended Coma cluster
galaxies, as following: A = NGC 4839, B = NGC 4827, C = NGC 4789. $10' \approx
 400$ kpc $h_{50}^{-1}$.}
\end{figure}
\begin{itemize}
\item 
Peripherally located sources of irregular extended radio emission.

\item
Steep radio spectra $\alpha = 1.0-1.5$.

\item 
Typical radio luminosities of $10^{41...42}$ erg s$^{-1}$.

\item
Radio relics are believed to be the remnants of radio lobes of radio
galaxies, where the former active galaxy has become inactive or has
moved away.

\item
But cluster radio relics have frequently no nearby possible parent
galaxy within an electron cooling-time travel distance.

\item
Further: The radio spectra of cluster radio relics do usually not show an
apparent cutoff, as the spectra of old remnants of radio galaxies do.

\item 
Thus the electron population is (re)accelerated there.

\item
Nine known examples: 0038-096 in A85, 0917+75 in A786, 1140+203 in
A1367, 1253+275 in Coma, 1712+64 in A2255, 1706+78 in A2256, 2006-56
and 2010-57 in A3667, 1401-33 in S753

\end{itemize}

\section{ A Simple Model}
\begin{figure}
\psfig{figure=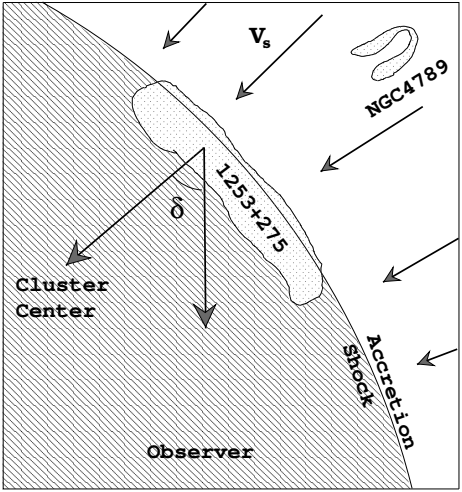,width=0.5\textwidth}
\vspace{0.8cm}
\caption[]{\label{fig:sketch}Sketch of the assumed geometry in the
Coma cluster.  The accretion flow is indicated by arrows. The shaded
region is the intra-cluster medium. The dotted area represents regions
of radio plasma. The line-of-sight and the line connecting the relic
and the cluster center are lying within the plane of the
figure. NGC4789 is placed upstream above the radio relic as a source
of relativistic electrons. Its line-of-sight velocity component points
away from the observer.  }
\end{figure}
The orientation of the magnetic fields of the infalling plasma is
assumed to be distributed randomly. The plasma is compressed at the
shock by the compression ratio $R$. This determines the spectral index
of the electron population. The high energy cutoff of the downstream
electron population is a function of the distance to the shock
plane. High resolution radio observations should therefore reveal a
flatter (steeper) spectrum at the outer (inner) edge of the relic,
which is closer (more distant) to the accretion shock. This is indeed
observed for the relics in Coma and A3667.  The overall radio spectrum
is a superposition of the spectra at different distances and is
composed by a flat low frequency spectrum, then a spectral index
steeper by $0.5$ due to the superposition of different cutoffs, and a
final cutoff. The observed spectral indices of relics belong to the
steeper region, expect in the case of the relic in A2256, which is not
due to a peripheral accretion shock but due to a massive cluster
merger. The shock compression ratio can be calculated\footnote{Since
particle acceleration is most efficient for oblique shocks (Jokipii
1987), patches with magnetic fields oriented parallel to the shock
plane will dominate the acceleration. The spatial diffusion
coefficient, which determines the efficiency of the acceleration
process, is only poorly understood in such complicated
circumstances. Thus, we assume a momentum independent coefficient,
because of the success of this simplification in the similarly
complicated case of SNR CR-acceleration (Biermann 1993).}
from the observed radio spectrum:
\begin{equation} \label{eq:R}
R = ({\alpha+1})/({\alpha-0.5})\,\, ,
\end{equation}
The ratio of break- to cutoff-frequency allows (if observed) a rough
estimate of the electron diffusion coefficient (En{\ss}lin et al. 1998a):
\begin{equation}
\label{eq:kappa2}
 \kappa_2 \approx {V_{\rm s}\,D}\,{R}^{-1}\, \sqrt{{\nu_{\rm
 break}}/{\nu_{\rm cut}}} < 3 \cdot 10^{30} \,{\rm cm^2\,s^{-1}} \,\,.
\end{equation}
The thickness $D \approx 0.1$ Mpc, and the other parameters are taken
for the relic 1253+275 in Coma. The thickness divided by the post
shock velocity $V_{\rm s}/R$ gives the age of the relic after it
passed the shock. It should be similar to the cooling time of
electrons visible at the spectral break.  Both ages agree roughly for
all known relics to be of the order of $\approx 10^8$ yr. The shock
efficiency which is necessary in order to power the radio emission is
$0.1...5\%$.

\section{ Radio Polarization}
\begin{figure}
\psfig{figure=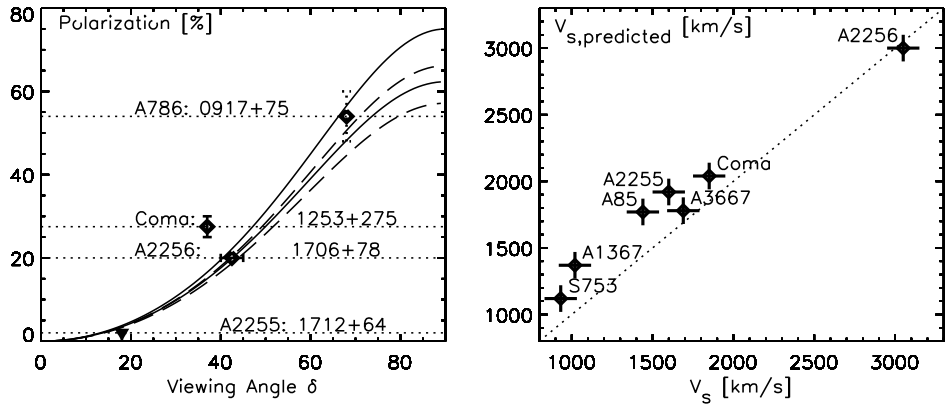,width=\textwidth}
\vspace{0.8cm}
\caption[]{\label{fig:pol}Polarization of synchrotron emission of
shocked magnetic fields as a function of the viewing angle
$\delta$. The curves are theoretical predictions for different radio
spectral indices and different magnetic field strength (En{\ss}lin et
al. 1998a). Observed polarizations of radio relics are plotted above
the viewing angle predicted from the accretion shock theory (1253+257,
1712+64), simulations of a cluster merger (1706+78, Roettiger et
al.~(1995)), or -- if no data are available -- at their best-fit
position (0917+75). The large observed projected radius of 0917+75
indicates a large viewing angle, consistent with the best-fit
position. Uncertainties in the angles are large.}

\caption[]{\label{fig:vs}A comparison of the shock velocity predicted
from simulations of structure formation $V_{\rm s, predicted}$
(Eq. \ref{eq:Vs}), and $V_{\rm s}$ estimated using the derived
compression ratio from the relic spectral index. The good agreement
demonstrates self-consistency of our simple model. Error bars are
large. In the case of A2256 the predicted shock velocity is derived by
Roettiger et al.~(1995) by fitting cluster merger scenarios to the
observed X-ray profiles.}

\end{figure}

The magnetic fields of the old radio lobe get compressed in the shock
and therefore aligned with the shock plane. If the shock is observed
inclined by a viewing angle $\delta$ the projected fields are
perpendicular to the projected shock plane. For relics with observed
radio polarization (Coma, A2256, A786) this is true. If the magnetic
fields are not dynamical important, the radio polarization of fields,
which were originally randomly oriented, can be estimated
(En{\ss}lin et al. 1998a). Since an independent very rough estimate of
the viewing angle $\delta$ is available from Eq. \ref{eq:rs} and the
projected observed radius, the expected correlation between
polarization and expected viewing angle can be checked (Fig. \ref{fig:pol}).

\section{ Probing Large-Scale Flows  }
Cluster radio relics allow to measure properties of the large-scale
flows.  An extrapolation of the cluster gas density by the usual
$\beta$-profile divided by the compression ration gives an estimate of
the density of the infalling gas: $n_{\rm e} \approx0.5...1.0\cdot
10^{-5}\,{\rm cm^{-3}}$.  The temperature jump of the shock is also
given by the compression ratio. Assuming that the temperature inside
the shock is half of the central cluster temperature, which is a
typical observed decrease, one gets for the temperature of the
infalling gas $0.5 ... 1$ keV. Also the predicted shock velocities
(Eq. \ref{eq:Vs}) can be checked: Using the observed parameters of the
shock the velocity $V_{\rm s}$ of the infalling matter can be
estimated. This is consistent with the prediced one ($V_{\rm
s,predicted}$) from cosmological structure formation (Fig. \ref{fig:vs}).

\section{ Conclusions}
\begin{itemize}
\item 
Giant shock waves of the cosmological large-scale motion of the on-going
structure formation are places of particle acceleration.

\item
The conditions for particle acceleration are ideal if an old radio lobe is
dragged into such a shock wave.

\item
{\it First observational evidence for these shock waves is presented}:
Cluster radio relics trace accretion shock fronts where the gas of
large-scale filaments flows into clusters of galaxies

\item
Observed properties of cluster radio relics can be explained if they
are understood as remnant radio plasma within cluster accretion
shocks.

\item
Properties of the large-scale flows can be measured.

\item
The long lasting outstanding problem of the energy supply of cluster
radio relics is solved by the accretion shock theory.

\end{itemize}
{\large \bf References}\\
%
Biermann P.L., 1993, A\&A { 271}, 649\\
Jokipii J.R., 1987, ApJ 313, 842\\
En{\ss}lin T.A., Biermann P.L., Kronberg P.P., Wu X.-P., 1997, ApJ
477, 560\\
En{\ss}lin T.A., Biermann P.L., Klein U., Kohle S., 1998a, A\&A 332,
395\\
En{\ss}lin T.A., Wang Y., Nath B.B., Biermann P.L., 1998b,
A\&A 333, L47\\
Giovannini G., Feretti L., Stanghellini C., 1991, A\&A { 252}, 528\\
Kang H., Rachen J.P., Biermann P.L., 1997, MNRAS 286, 257\\
Roettiger K., Burns J.O., Pinkney J., 1995, ApJ 453, 634\\

\end{document}